\documentstyle[pra,aps,epsf]{revtex}
\begin{document}
\newcommand{\bfsigma}{\mbox{\boldmath $\sigma$}}
\draft
\title{Realistic effective interactions for halo nuclei }

\author{T.T.S. Kuo${}^1$, H. M\"uther${}^{2}$, and K. Amir-Azimi-Nili${}^2$}
\address{${}^1$Department of Physics, SUNY-Stony Brook\\
Stony Brook, New York 11794 USA\\
${}^2$ Institut f\"ur Theoretische Physik, Universit\"at T\"ubingen\\
D-72076 T\"ubingen, Germany}
\date{\today}
\maketitle

\begin{abstract}
We study halo nuclei using a two-frequency shell-model approach employing
wave functions of two different oscillator constants $\hbar\omega_{in}$
and $\hbar\omega_{out}$, the former for the inner orbits and the latter for
the halo (outer) orbits. An initial application
has been made for the halo nuclei $^6$He and $^6$Li,
with  $0s_{1/2}$ taken as the inner and ( $0p_{3/2}$, $0p_{1/2}$)
as the halo orbits. Starting from the Paris NN interaction, we have derived
a G-matrix folded-diagram effective interaction for this two-frequency
model space, using an essentially exact treatment for the Pauli exclusion
operator for the G-matrix. While keeping $\hbar\omega_{in}$ fixed, we have
performed calculations with different choices for
$\hbar\omega_{out}$, treating it as a variation parameter.
For $\hbar\omega_{in}=19.7 MeV$ and  $\hbar\omega_{out}=8.2 MeV$, our
calculated valence energies for $^6$He and $^6$Li are  -2.77
and -3.55 Mev, respectively, both in good agreement with experiments.
The importance of certain three-body-force diagrams is discussed.

\end{abstract}


\newpage

\section{ Introduction }

Since several years ago, radioactive-beam nuclear physics has been rapidly
developing and has attracted the attention of many nuclear physicists
\cite{vaag93,mue93,boyd94,bert89,ren95,naka94,otsu96,orma}.
With radioactive-beam accelerators, one is now able to measure
the properties of a large number of drip-line nuclei, which are
far from the nuclear stability valley. These nuclei are generally barely bound
(or unbound) and their structure is typically that of a tightly bound
inner core with a few outer nucleons which are loosely attached to the core.
 For example,
the halo nucleus $^6$He is presumably made of a $^4$He core with a pair
of loosely-attached outer nucleons. The outer nucleons are
spatially extended and have a large rms radius, forming a "halo" cloud rimming
the nucleus as illustrated in Fig. 1. Such nuclei are usually called
 "halo nuclei".  Vaagan et al. \cite{vaag93} have suggested a more distinctive
name for them, namely  "Borromean nuclei", as a number of halo nuclei,
such as $^{11}$Li, are believed to form a special type of bound three-cluster
 system.

Studies of halo nuclei, or nuclei far off stability, are of interest
 in many ways.
In astrophysics, the location of the r-process (rapid proton and/or neutron
absorption process) path plays a crucial role for nucleosynthesis.
The r-process consists of a chain of nuclear reactions, involving a number
of intermediate nuclei which are far off
stability. To locate the path of the r-process accurately, it is necessary
 to explore, as much as we can, the various properties of such nuclei.
With radioactive beam facilities, the cross section for certain such nuclear
reactions important to nucleosynthesis can now be directly measured.
For example, the cross section for $^8$Li$(\alpha,n)^{11}$B has been measured
at the radioactive beam facility of RIKEN \cite{boyd92}.

Halo nuclei may play also an important role for the synthesis of super-heavy
elements, which has been pursued by many physicists.
The use of halo nuclei, which are already far off stability, as projectiles
may provide a new and possibly more efficient avenue for producing
super-heavy elements.  Also from a theoretical viewpoint, the study of
halo nuclei is of high interest. The empirical nuclear mass
formula is, so far, based on a
fit to the observed properties of stable nuclei. We need information about
halo nuclei to generalize the mass formula to regions far off nuclear stability.

It seems to be still rather uncertain about what would be an
appropriate and convenient theoretical framework
for the description of halo nuclei. As will be discussed in Section 2, the
ordinary nuclear shell model may not be suitable for halo nuclei.
We would like to suggest a two-frequency shell model, using single particle
(sp) wave functions of two different oscillator constants $\hbar\omega_{in}$
and  $\hbar\omega_{out}$, the former for the inner orbits while the latter
for the halo (outer) orbits.

It is our great pleasure and honor to dedicate this short article to
 Professor G.E. Brown, or Gerry as known to all
his friends, colleagues, students and tennis partners, on the occasion
of his 70th birthday. Gerry has pioneered the G-matrix approach for
microscopically deriving the
nucleon-nucleon effective interaction ($V_{eff}$) in nuclei.
\cite{brow67,brok67,kuob66}

This approach has turned out to be quite successful, for ordinary nuclei.
For example, the CalTech group has recently carried out shell-model
Monte-Carlo calculations \cite{lang95} using G-matrix effective
interactions, with  results in remarkably good
agreement with experiments  over a wide range of nuclei.
 In this article, we would like to discuss an extension of  the above
G-matrix approach to halo nuclei, and report some preliminary results.

In the following, we shall first discuss our two-frequency model space and
G-matrix together with some calculational details in Section 2.
The importance of certain three-body diagrams to the effective interaction
will be addressed. In Section 3, we shall present some preliminary
results for halo nuclei $^6$He and $^6$Li. A summary and discussion will be
presented in Section 4.

\section{Two-frequency model space and G-matrix}

The rms radius for a nucleon in a harmonic-oscillator shell-model orbit
$\phi _{nl}$ is given as
\begin{equation}
r_{rms}(nl)=(2n+l+\frac{3}{2})^{1/2}b;\qquad \mbox{with}\qquad
b\equiv (\hbar/m\omega)^{1/2}.
\end{equation}
To fit the observed nuclear radii, there is an empirical formula
\cite{bert72} for choosing the oscillator parameter $\hbar\omega$,
namely
\begin{equation}
\hbar\omega=45A^{-1/3}-25A^{-2/3}
\end{equation}
where A is the nuclear mass number. Using this formula, the empirical
value of $\hbar\omega$ is approximately 17.2 MeV
for $^{6}$He, corresponding to a length parameter b= 1.55 fm.

In ordinary shell model calculations, it is customary to use the above
empirical formula to determine the $\hbar\omega$ value used for
the shell-model wave functions. For example,
we may use $\hbar\omega$=14 MeV (b= 1.72 fm) for most of the sd-shell nuclei.
The situation for halo nuclei is, however, different. For instance, the
empirical rms radius for $^6$He is about 2.5 fm \cite{vaag93}, which corresponds
to a b value much larger than the above value of 1.55 fm.
For halo nuclei such as $^6$He, we need to have wave
functions with larger spatial distribution, compared with the ordinary
shell-model wave functions.

Let us also look at the binding energies of some nuclei in the vicinity of
$^6He$. The observed values \cite{wall95} are: 28.29 ($^4$He), 27.40
($^5$He),
 26.32 ($^5$Li), and  29.26 ($^6$He), in MeV. We see that the two-neutron
separation energy for
$^6$He is only about 1 MeV, and the single-nucleon separation energies for
$^5$He and $^5$Li are both negative (unbound).
It is clearly seen that, comparing to ordinary nuclei, $^6$He is a very
"special" breed. It appears to have a tightly-bound $^4$He core with a
pair of very loosely bound outer nucleons. For this situation, it would be
very difficult for the ordinary shell model, where all the wave functions
have a common $\hbar\omega$ value, to give an adequate description for
the nuclear wave function; to have a good
description for the $^4$He core we need to use shell-model wave functions
with a small b value, $\sim 1.45 fm$, which is much smaller than the
b value needed for a good description for the outer nucleons.
In fact, to avoid this serious difficulty, many physicists
(see [1] and references quoted therein)
have abandoned the ordinary shell-model approach, and employed instead
the method of hyperspherical harmonics \cite{vaag93}
for the description of halo nuclei.

We would like to consider a two-frequency shell-model (TFSM) approach for halo
nuclei, as described below. In recent years, advance has been made for
performing realistic nuclear structure
calculations starting from a fundamental nucleon-nucleon (NN) potential,
such as the Paris potential\cite{paris}. The effects of correlations,
which are not explicitly included in the model space, are taken into
account by deriving the effective hamiltonian for this model space from
the realistic NN interaction. For the hyperspherical harmonics
method, there is still the restriction that one  has to employ
a phenomenological effective interaction, such as a Gaussian interaction.
 This seems to be a drawback. To our knowledge, it is not
yet clear how to incorporate realistic NN interactions
into the hyperspherical harmonics method. In contrast, for our TFSM approach
it seems to be straightforward to perform microscopic calculations for halo
nuclei, starting from realistic NN interactions.
Otsuka and his coworkers \cite{otsu96} have proposed a variational shell-model
approach for treating halo nuclei, and have carried out extensive calculations
using the Skyrme effective interactions VSM and SIII.

Let us use the p-shell nuclei $^6$He and $^6$Li
to illustrate our approach. It is reasonable to consider the "core" of these
nuclei as an ordinary alpha particle. Hence we take in the present
work the $0s_{1/2}$ orbit
as an inner orbit with oscillator constant $\hbar\omega_{in}$=19.7 MeV,
corresponding a length parameter $b_{in}$=1.45 fm. The outer, or halo, nucleons
of these nuclei are spatially extended, and for them we use
the halo orbits $0p_{3/2}$ and $0p_{1/2}$ with oscillator constant
$\hbar\omega_{out}$, which is treated as a variation parameter.
In this way, our sp  model space may be written as
\begin{equation}
P_{sp}=$\{$ \phi^{in},d^{in};\phi^{out},d^{out}$\}$
\end{equation}
with the total dimension of $P_{sp}$ being $d=(d^{in}+d^{out})$.
The superscripts
$in$ and $out$ refer to the inner and outer orbits, respectively.
Clearly one needs to have all the sp wave functions
$\{\phi^{in},\phi^{out}\}$ be orthonormal to each other, and one may be
concerned as to how can one fulfill this condition, as $\hbar\omega_{in}$
is  in general not equal to $\hbar\omega_{out}$. For a small model
space, the orthogonality condition
is easily satisfied because of parity and angular momentum conservation.
For a general case, we need to use
a common $\hbar\omega$ value for all the orbits with the same
$l$ and $j$ values. Thus, for our present $^6$He and $^6$Li case,
we use $\hbar\omega_{in}$ not only for the $0s_{1/2}$ orbit but also for all
the other $s_{1/2}$ orbits, in order to fulfill the above orthonormality
condition. (Note that the $ns_{1/2}$ orbits, with $n>0$, are needed for the
intermediate states in the evaluation of the effective interaction as
discussed below.)


For a general model space, the model-space effective interaction
$V_{eff}$ can be
derived formally from a folded-diagram method \cite{kuos}. Usually this
derivation is carried out in the following three steps.
First, we  calculate the  model-space Brueckner G-matrix
 defined by the integral equation \cite{kkko,muet92}
\begin{equation}
G(\omega)=V+V Q_2\frac{1}
{\omega-Q_2TQ_2}Q_2G(\omega),
\end{equation}
where $\omega$ is an energy variable. $Q_2$ is a two-body Pauli exclusion
operator, and its treatment is very important as we shall later discuss.
$T$ is the two-nucleon kinetic energy. Note that our G-matrix has
orthogonalized plane-wave intermediate states.
In the second step, the irreducible vertex function $\hat Q-box$ is
calculated from the above G-matrix. Finally the energy-independent
effective interaction
is given by the folded-diagram series
\begin{equation}
V_{eff}= \hat{Q} - \hat{Q}^{'}\int\hat{Q}
+ \hat{Q}^{'}\int\hat{Q}\int\hat{Q} - \hat{Q}^{'}\int\hat{Q}\int\hat{Q}
\int\hat{Q}\cdots,
\end{equation}
where $\int$ denotes a generalized fold, and $\hat{Q}^{'}$ and $\hat{Q}$
represent the $\hat Q-box$ \cite{kuos}.

The above formalism is general, and is applicable to either the situation
with a single oscillator frequency as in earlier shell-model calculations,
or the present situation with two oscillator frequencies.
For the latter, the calculation of the G-matrix is  more complicated.
The exact solution \cite{kkko} of the G-matrix of
Eq.(4) is
\begin{equation}
G=G_F+\Delta G
\end{equation}
where the "free" G-matrix is
\begin{equation}
G_{F}(\omega)=V+V\frac{1}{\omega-T}G_{F}(\omega),
\end{equation}
and  the Pauli correction term ${\Delta G}$ is given by
\begin{equation}
\Delta G(\omega)=-G_F(\omega)\frac{1}{e}P_2\frac{1}
{P_2[1/e+(1/e)G_F(\omega)(1/e)]P_2}
P_2\frac{1}{e}G_F(\omega)
\end{equation}
where $e=\omega-T$. The projection operator $P_2$ is defined as (1-$Q_2$).

The basic ingredient for calculating the above G-matrix is the matrix elements
of $G_F$, the free G-matrix, within the $P_2$ space. This space is composed
of, however, wave functions of two frequencies, $\hbar\omega_{in}$ and
$\hbar\omega_{out}$. This poses a technical difficulty because
transformations from  two-particle states in the c.m.
coordinates to those  in the laboratory coordinates are not as easy to
perform as in the case of one oscillator frequency.
We have adopted an expansion
procedure to surmount this difficulty, namely expanding the
oscillator wave functions with $\hbar\omega_{in}$ in terms of those with
$\hbar\omega_{out}$, or $vice ~versa$. When $\hbar\omega_{in}$ and
$\hbar\omega_{out}$ are close to each other, this procedure is relatively
easy to carry out. But when they are significantly different,
the two-frequency
G-matrix is considerably more complicated to calculate than the one-frequency
one.

We write the projection operator $Q_2$ as
\begin{equation}
Q_2=\sum_{all~ab} Q(ab)\vert ab\rangle \langle  ab \vert,
\end{equation}
and define Q(ab) as
\begin{equation}
$$Q(ab)=\cases{0,&if $b\leq n_1,a\leq n_3$;\cr
               0,&if $b\leq n_2,a\leq n_2$;\cr
               0,&if $b\leq n_3,a\leq n_1$;\cr
               1,&otherwise.\cr}$$
\end{equation}
As shown in Fig. 2a, the boundary of Q(ab) is specified by the orbital
numbers ($n_1,n_2,n_3$). We denote the shell model orbits by numerals,
starting from the bottom of the oscillator well, 1 for orbit $0s_{1/2}$,
2 for $0p_{3/2}$,...7 for  $0f_{7/2}$ and so on. $n_1$ denotes the
highest orbit of the closed core (Fermi sea). $n_2$ denotes the highest orbit
of the chosen model space. We consider here $^6$He and $^6$Li with $^4$He
treated as a closed core, thus
we have $n_1$=1. Suppose we use a model space including all the 6 orbits
in the s, p and sd shells. Then for this case $n_2$=6.
In principle one should take $n_3=\infty$ \cite{kkko}. In practice, this is not
feasible, and one can only use a large $n_3$ determined by an empirical
procedure. Namely, we perform calculations with increasing values for $n_3$
until  numerical results become stable.
As illustrated in Table 1, we see that there is a convergence behavior and
for our present calculation a choice of
$n_3$=21 appears to be adequate. (This $n_3$ value will be
used in our present work.)

There is an important point about whether we use a large $n_3$ or not.
As illustrated in the diagram of
Fig. 3 (upper part), we used railed lines to denote nucleons outside 
the model space
while for those inside  we use  bare lines.
This diagram  has one line outside the model space, and hence is a legitimate
G-matrix diagram, even if the state $b$ refers to a single-particle
state below the Fermi energy (i.e. $b\leq n_{1}$).
We are, however, excluding this diagram from the G-matrix
when we use a large $n_3$. How can we do this?

There are 3-body components in the effective interaction.
Consider the example of the diagram of Fig. 3a (lower part), which represents a
3-body-force (irreducible)
diagram, where $j$ denotes, for example, a particle in orbit $0s_{1/2}$ and
$a$ represents an orbit outside the model
space. (We now draw diagrams with respect to the bare vacuum.)
Note that the exchange
diagram of (3a) is just diagram (3b) with a minus sign.
Diagram (3b) is, however, equivalent to the diagram of Fig. 3 (upper part). 
Hence the
diagrams of the type shown in Fig. 3 (upper part) are 
exactly canceled by the corresponding
3-body-force diagrams of the type shown in Fig. 3 (lower part).
In other words,  such 3-body-force diagrams can be incorporated into the
G-matrix and in so doing they cancel the G-matrix diagrams
of the type shown by Fig. 3 (upper part). Consequently 
we should use a large $n_3$.

It is common to employ a simpler Pauli
operator \cite{barr} (and references quoted therein), defined by " $Q(ab)=1,
 ~if ~(\rho_{a}+\rho_{b}) \leq N_0 ;~  =0,$ otherwise", with
$\rho_a \equiv 2n_a+l_a$
and similarly for $\rho_b$. Such a Pauli operator has the shape
shown in Fig. (2b). Here $N_0$ is typically a small number, such as 2 for the
p-shell nuclei. A main advantage of this choice is its convenience for
numerical calculation. However,  this choice for Q corresponds to a small
$n_3$, and hence  the 3-body-force diagrams as shown by
Fig. 3 (lower part) are not yet taken into 
account by the small-$n_3$ G-matrix;
they have to be calculated separately. In contrast, we use a large
$n_3$ in our present work, and
hence the effect of such 3-body-force diagrams is already absorbed
in our G-matrix. Using a small-$n_3$
G-matrix alone may over-estimate the nuclear binding energy, as the effect
of such 3-body-force diagrams is usually repulsive.

\section{ Application to $^6$He and $^6$Li }

 Experimental data about $^6$He and $^6$Li are rather scarce.
 The valence energies of these nuclei, defined as
\begin{equation}
 E_{val}(^6\mbox{He})=-[BE(^6\mbox{He})+BE(^4\mbox{He})-2\times BE(^5
   \mbox{He})]
\end{equation}
and
\begin{equation}
 E_{val}(^6\mbox{Li})=-[BE(^6\mbox{Li})+BE(^4\mbox{He})- BE(^5
   \mbox{He})-BE(^5\mbox{Li})],
\end{equation}
 are nevertheless  well known. From the empirical binding energies (BE)
\cite{wall95}, the valence energy
for $^6$He is found as -2.77 MeV and that for $^6$Li is -3.83 MeV.

We want to calculate these quantities using a TFSM approach. For the
inner orbit $0s_{1/2}$ we use $b_{in}$=1.45 fm. For the outer
orbits $0p_{3/2}$ and $0p_{1/2}$ we use a sequence of $b_{out}$
values, ranging from 1.45 to 2.25 fm. The G-matrix is first
calculated using a Pauli operator specified by $(n_1,n_2,n_3)$=(1,6,21),
starting from the Paris
potential. Then, following closely the $\hat Q-box$ folded-diagram
procedure of Shurpin et al. \cite{shur83}, we derive the matrix
elements of the effective interaction $V_{eff}$ for the p-shell.
(In the $\hat Q-box$ we have included diagrams first- and second-order
in G. This includes the diagram of second order in $G$ with an
intermediate state of two nucleons in $1s0d$ shell. To avoid
overcounting we have to use $n_{2}=6$ in the Pauli operator for the
G-matrix equation.)  Our results for the matrix elements of
$V_{eff}$ are listed in Table 2.
As expected, these matrix elements are seen to become
generally weaker as $b_{out}$ increases, a reflection that the mean
distance between the interacting valence nucleons is larger.
The empirical (6-16) matrix elements of Cohen and Kurath \cite{coku}
are also listed for comparison. For some cases, the agreement between
our calcualated matrix elements and their values becomes improved
when we use a $b_{out}$ value larger than 1.45 fm.

To obtain the valence energies, we just diagonalize the two-particle
matrix $(H_0+V_{eff})$ within the p-shell, using the matrix elements of Table
2.  $H_0$ is the sp Hamiltonian.
For the valence-energy calculation, the single-particle energy for
$p_{3/2}$ is defined
as zero. But that for $p_{1/2}$ is uncertain, both experimentally and
theoretically. For an initial calculation, we have fixed it as 10 MeV.
(Using an empirical formula given in Bohr and Mottelson's book \cite{bomo},
the $p_{1/2}-p_{1/3}$ spin-orbit splitting is estimated to be $\sim 10.2MeV$.)
The valence energy for $^6$He is then just the
lowest eigenvalue of the T=1, J=0 secular matrix, and that of
 the T=0, J=1 matrix gives the valence energy for $^6$Li.
Our results for these energies are displayed in the lower part of Fig.~4,
calculated as a function of $b_{out}$.
At $b_{out}\simeq $ 1.45 fm, the calculated values are too large.
At  $b_{out}=$ 2.25 fm, our results are -2.77 and -3.55 MeV for $^6$He
and $^6$Li, respectively, both in reasonable agreement with the corresponding
empirical values -2.76 and -3.83 MeV.

We have also calculated the single-particle separation energy of
$^5$He, defined as
$E_{sp}=BE(^5$He$)-BE(^4$He$)$. Our calculation was done using a one-body
$\hat Q-box$ consisting of the three 1-body  diagrams
(first- and second-order in G) of Ref.\cite{shur83} and then summing up
the  folded-diagram series consisting of the one-body $\hat Q-box$.
These results are presented in the upper part of
Fig.~4. There is an improvement of our results as $b_{out}$ increases.
However, our calculated separation energy  is still too large at
$b_{out}\simeq$2.25 fm.
It is of interest to note that there seems to be a saturation
point for $E_{sp}$ at  $b_{out}\simeq$2.25 fm. We recall that at this
$b_{out}$
our results for $E_{val}$ are also in good agreement with experiments.
Assuming a pure two-frequency $s^4p^2$ wave function with
$b_{in}$=1.45 and $b_{out}$=2.25 fm , the rms radius
for $^6$He is obtained, using Eq.(1), as 2.37 fm, which is also in reasonable
agreement with the empirical value of $\sim 2.5 fm$ \cite{vaag93}.

Encouraged by the above preliminary results, we have begun to tackle
a more difficult problem, to calculate the total
binding energies of $^4$He, $^5$He and $^6$He using our present two-frequency
approach. Here we employ a two-shell ( s and p) model space, treating
all nucleons of the nucleus as active. For s-shell we use fixed length
parameter $b_{in}$ = 1.45 fm. For p-shell we use a varied $b_{out}$.
The components of
a spurious c.m. motion, which are contained in such a model space
including valence orbits of 2 major shells are removed using the
technique of Gloeckner and Lawson \cite{law}.
Our preliminary results indicate that for
 $^4$He an energy minimum seems to be at $ b_{out}\simeq 1.45 fm$, while
 for $^6$He the minimum seems to be located at $ b_{out}\simeq 2.25fm$.
Our calculated total
binding energies have turned out to be too small, for both nuclei,
compared with experiments. This is perhaps a common problem of nuclear
structure calculations with no-core employing realistic NN
interactions. Further studies
are being pursued, and we hope to report our results before long.

\section{Summary and discussion}

 Halo nuclei have a special nuclear structure; they have typically
a tightly bound core with a few halo nucleons loosely attached to the core.
One would encounter a lot of difficulties, in describing such nuclei
using the conventional shell model where the basis functions all have
the same $\hbar\omega$ value. For example, the wave functions with
$b_{in}$=1.45 fm would give a good description for the alpha-particle
core of $^6$He, but they are spatially much too compressed to describe the
halo neutrons of $^6$He. If one insists to use the compressed wave functions
also for the halo nucleons, then one needs to use a very large active
space (such as a space with s, p, sd and pf shells for the case of $^6$He).
 This would be very tedious as well as not economical. A much simpler solution
is to allow for a two-frequency model space, where one uses $b_{in}$
for the core and $b_{out}$ for the halo part of the nucleus
as described in the present work. We have obtained some
preliminary results using this approach. Our results seem to indicate
that this two-frequency method is a feasible and promising approach
for halo nuclei.

What effective interactions should one use in a two-frequency shell-model
calculation? It would be very difficult to determine them by an empirical
best-fit procedure, which has been successfully used for one-frequency
shell-model calculations \cite{wild}. This is  because now the
length parameter $b_{out}$ is a variable and the matrix elements of
the effective interaction are dependent on it. A basic and rather "ambitious"
solution to this problem would be to derive the effective interaction
directly from a fundamental NN interaction. We have carried out some
preliminary studies in this direction. We first calculate the two-frequency
G-matrix, with an accurate treatment for its two-frequency Pauli
exclusion operator. Then the effective interaction is obtained via a standard
folded-diagram method. Our calculated valence energies for $^6$He
and $^6$Li are in satisfactory agreement with experiments. As indicated by
our preliminary results,
$^4$He seems to favor a one-frequency ($b_{in}$=$b_{out}$=1.45 fm) shell model,
while a two-frequency ($b_{in}$=1.45, $b_{out}$=2.25 fm) one is perhaps needed
for $^6$He, $^6$Li and $^5$He.

\vskip 0.2cm
{\bf Acknowledgements}
\vskip 0.2cm
 Part of this work was performed when TTSK was a summer visitor at
T\"ubingen. He would like to thank Professors A. Faessler and H. M\"uther
for their warm hospitality during his stay at T\"ubingen. This work is
supported in part by "Graduiertenkolleg Struktur und Wechselwirkung von
Hadronen und Kernen" (DFG Mu 705/3) and by the USDOE Grant DE-FG02-88ER40388.

\newpage

\begin{table}[hb]
\begin{center}
\begin{tabular}{ l l l l l }
\hline\hline

 &  &        & $\omega$ [MeV] & \\

 $  a \ b  \ c  \ d  \ T\ J$ \ &$ n_{3}$ \ & -5 \ & -10
\ & -20
\\ \hline

  1 1 1 1 0 1 & 3& -16.048& -15.424& -14.453\\
  1 1 1 1 0 1 & 6& -15.295& -14.706&-13.842\\
  1 1 1 1 0 1 & 15& -15.238&-14.637&-13.767 \\
  1 1 1 1 0 1 & 21& -15.238&-14.637&-13.767 \\
\\
  2 3 2 3 0 2 & 3& -6.138& -5.691& -5.132\\
  2 3 2 3 0 2 & 6& -5.635& -5.339& -4.921\\
  2 3 2 3 0 2 & 15& -5.563& -5.277& -4.877\\
  2 3 2 3 0 2 & 21& -5.563& -5.277& -4.877\\
\\
  1 1 1 1 1 0 & 3& -9.456& -9.360& -9.213\\
  1 1 1 1 1 0 & 6& -9.151& -9.093& -8.995\\
  1 1 1 1 1 0 & 15& -9.140& -9.083& -8.986\\
  1 1 1 1 1 0 & 21& -9.140& -9.083& -8.986\\

\\ \hline\hline
\end{tabular}
\end{center}
\end{table}
Table 1. Dependence of the two-frequency G-matrix on the choice of $n_{3}$.
Listed are the matrix element $\langle abTJ \vert G(\omega) \vert cdTJ
\rangle$, in MeV, calculated with the Paris potential.
 The orbits 1,2,3 represent,
respectively, $0s_{1/2},0p_{3/2},0p_{1/2}$. We have used
$b_{in}$= 1.45  and $b_{out}$= 2.0 fm for the length parameters,
and $n_1=1$ and $n_2=3$ for the exclusion operator.

\newpage
\begin{table}[hb]
\begin{center}
\begin{tabular}{ l l l l l l}
\hline\hline

 &  &        & $b_{out}$ [fm] & & CK \\

 $  a \ b  \ c  \ d  \ T\ J$ \ &1.45 \ & 1.75 \ & 2.00 \ & 2.25&
\\ \hline

  2 2 2 2 0 1 & -1.76 & -2.15 & -2.28 & -2.24 & -3.14\\
  2 2 2 3 0 1 &  5.64 &  4.69 &  3.89 &  3.18 &  4.02\\
  2 2 3 3 0 1 &  2.55 &  2.24 &  2.13 &  2.03 &  1.09\\
  2 3 2 2 0 1 &  5.81 &  4.64 &  3.78 &  3.05 &  4.02\\
  2 3 2 3 0 1 & -7.63 & -6.75 & -5.96 & -5.15 &  -6.54\\
  2 3 3 3 0 1 &  3.12 &  1.72 &  0.83 &  0.20 &  1.39\\
  3 3 2 2 0 1 &  2.72 &  2.20 &  2.01 &  1.87 &  1.09\\
  3 3 2 3 0 1 &  3.11 &  1.82 &  1.08 &  0.56 &  1.39\\
  3 3 3 3 0 1 & -3.13 & -2.80 & -2.61 & -2.37 &  -4.26\\
\\
  2 2 2 2 1 0 & -3.43 & -3.28 & -2.92 & -2.51 &  -2.74\\
  2 2 3 3 1 0 & -5.21 & -3.88 & -3.02 & -2.38 &  -5.32\\
  3 3 2 2 1 0 & -5.41 & -3.86 & -2.99 & -2.36 &  -5.32\\
  3 3 3 3 1 0 &  0.44 & -0.45 & -0.74 & -0.81 &   0.34\\

\\ \hline\hline
\end{tabular}
\end{center}
\end{table}
Table 2. Some p-shell effective interactions calculated with various
$b_{out}$ values.
Listed are the matrix element $\langle abTJ \vert V_{eff} \vert cdTJ
\rangle$, calculated from the Paris potential.
 The orbits 2,3 represent, respectively, $0p_{3/2},0p_{1/2}$. We have used
$b_{in}$= 1.45 fm and bare G-matrix calculated with $(n_1,n_2,n_3)$=
(1,6,21). The empirical (6-16) matrix elements of Cohen and Kurath
\cite{coku} are listed under column CK.

\newpage






\begin{figure}[h]
\epsfysize=5cm
\begin{center}
\makebox[16.6cm][c]{\epsfbox{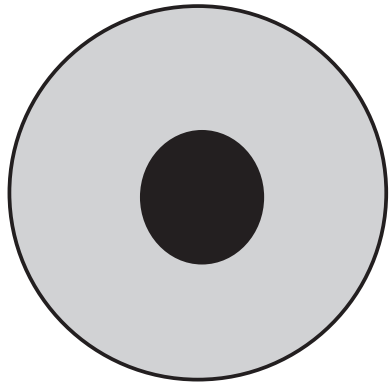}}
\end{center}
\vspace{1cm}
\caption{A halo nucleus.}
\end{figure}
\vspace{4cm}
\begin{figure}[h]
\epsfxsize=16.6cm
\begin{center}
\makebox[16.6cm][c]{\epsfbox{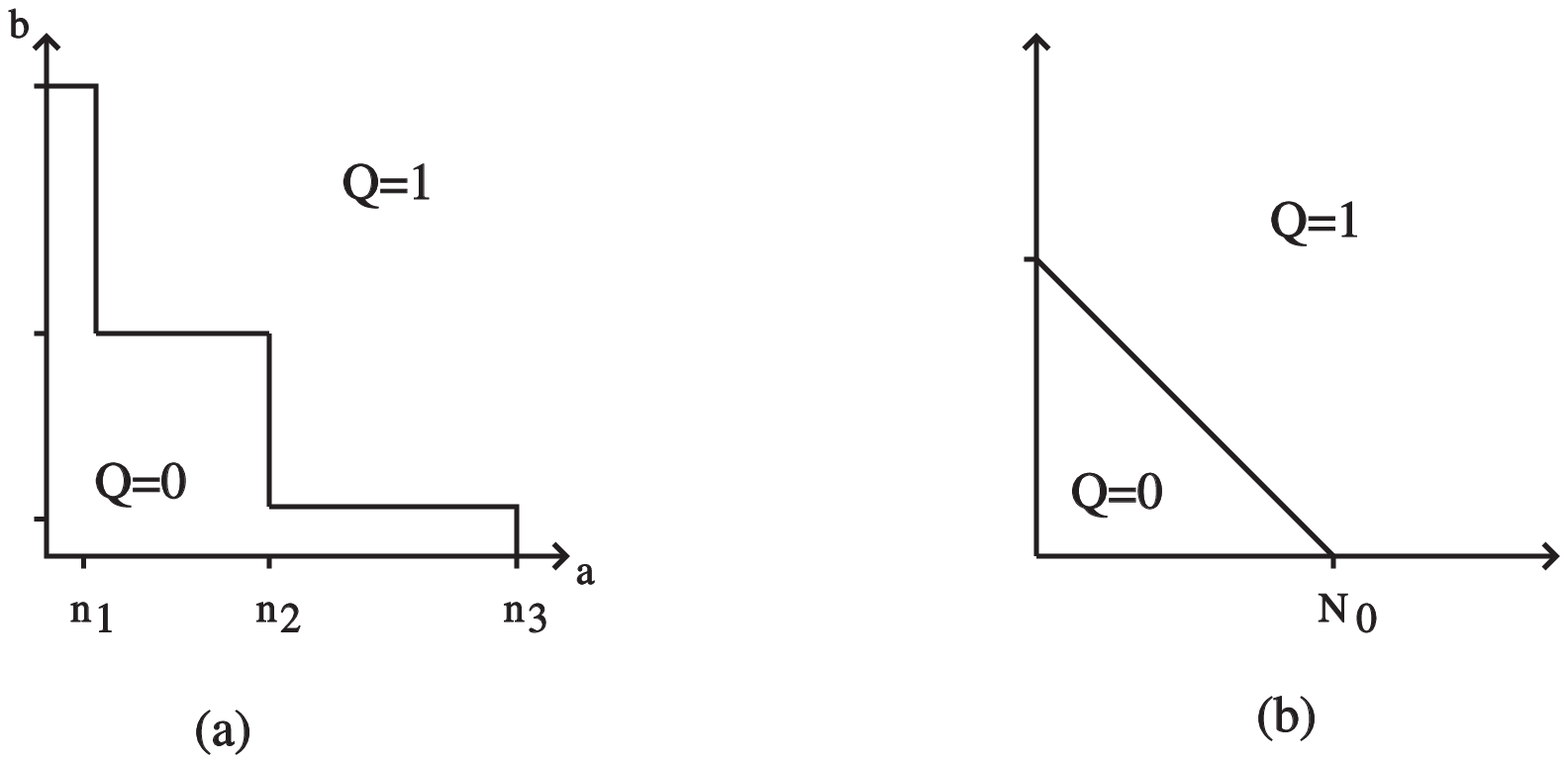}}
\end{center}
\caption{Pauli exclusion operator $Q_2$.}
\end{figure}

\begin{figure}[h]
\epsfysize=19cm
\begin{center}
\makebox[16.6cm][c]{\epsfbox{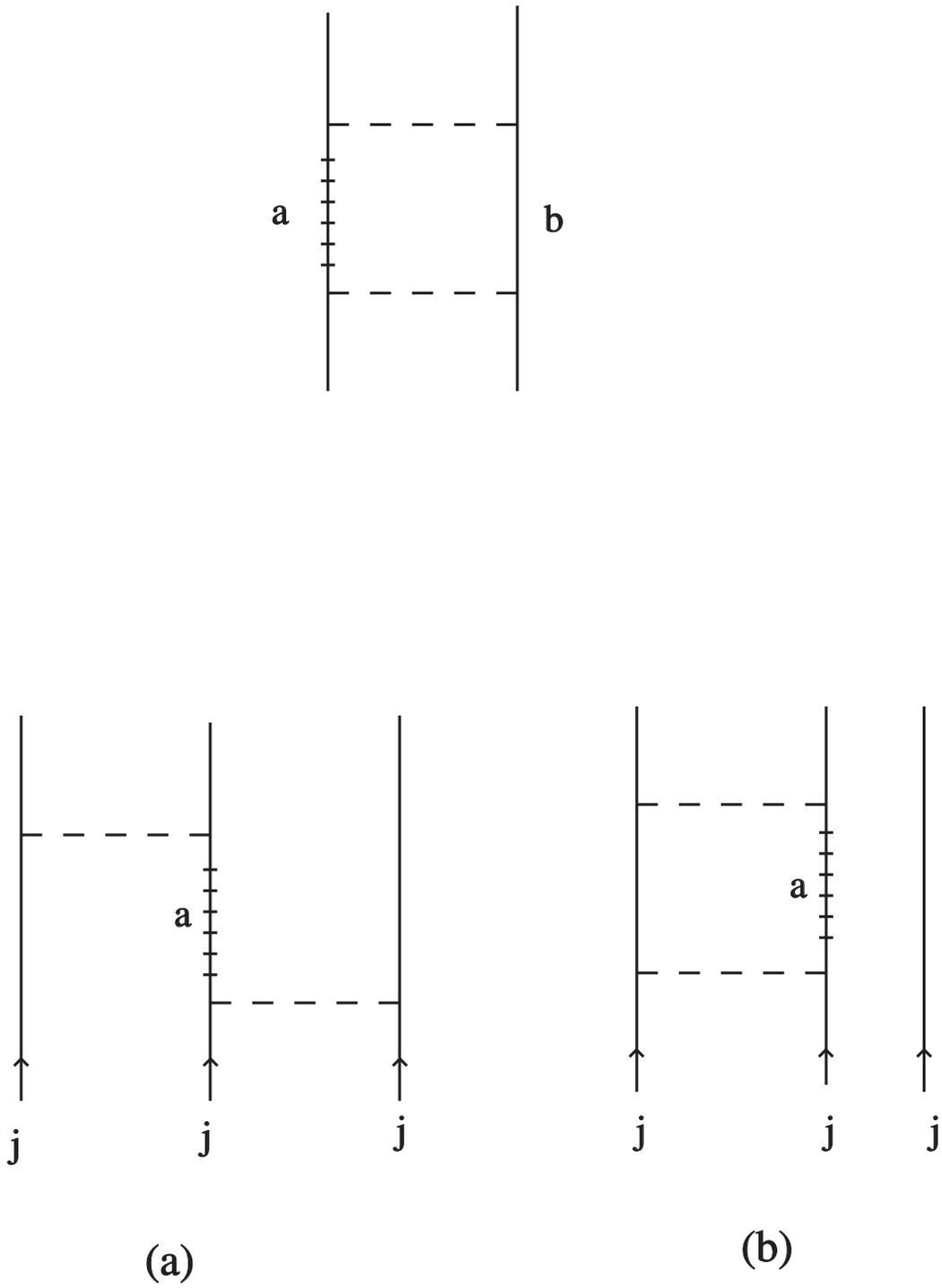}}
\end{center}
\vspace{2cm}
\caption{A second-order diagram belonging to the G-matrix (upper part) and 
Three-body-force diagrams (lower part).}
\end{figure}

\begin{figure}[h]
\epsfysize=22cm
\begin{center}
\makebox[16.6cm][c]{\epsfbox{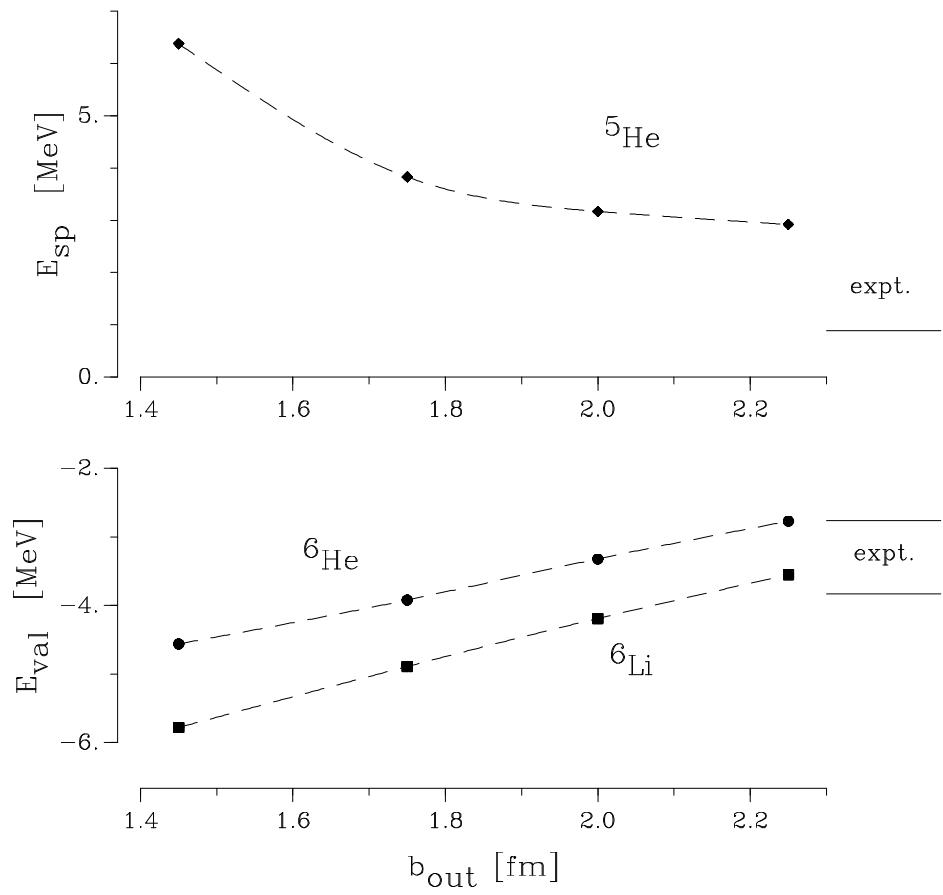}}
\end{center}
\vspace{-7cm}
\caption{ The lower part of this figure shows the valence energy $E_{val}$
for $^6$He and $^6$Li as a function of $b_{out}$, while the upper part
displays the single-particle separation energy $E_{sp}$ for $^5$He
as a function  of $b_{out}$.}
\end{figure}

\end{document}